\documentclass[12pt]{article}
\usepackage{graphicx}

\begin{document}

\pagestyle{myheadings}
\markright{Multiscale Trend Analysis}

\begin{titlepage}
\title{Multiscale Trend Analysis}

\author{Ilya Zaliapin \thanks{
Institute of Geophysics and Planetary Physics, 
University of California, Los Angeles,
CA 90095-1567, USA and
International Institute of Earthquake 
Prediction Theory and Mathematical Geophysics, 
Russian Academy of Sciences,
Moscow, Russia,
E-mail: zal@ess.ucla.edu
Phone: +10-310-8256115,
Fax: +10-310-2063051,
corresponding author.
}, \\
Andrei Gabrielov \thanks{
Departments of Mathematics and Earth and Atmospheric
Sciences, Purdue University, W.Lafayette, IN 47907-2067, 
USA. E-mail: agabriel@math.purdue.edu
}, and \\
Vladimir Keilis-Borok\thanks{
Institute of Geophysics and Planetary Physics and
Department of Earth and Space Sciences,
University of California, Los Angeles, 
CA 90095-1567, USA,
and 
International Institute of Earthquake Prediction Theory and
Mathematical Geophysics, Russian Academy of Sciences,
Moscow, Russia,
E-mail: vkb@ess.ucla.edu}} 

\bigskip
\date{May 04, 2003}

\maketitle
\end{titlepage}

\begin {abstract}
This paper introduces a multiscale analysis 
based on optimal piecewise linear approximations 
of time series.
An optimality criterion is formulated 
and on its base a computationally effective
algorithm is constructed for decomposition 
of a time series into a hierarchy of trends 
(local linear approximations) at different 
scales. 
The top of the hierarchy is the global linear
approximation over the whole observational
interval, the bottom is the original time 
series.
Each internal level of the hierarchy corresponds 
to a piecewise linear approximation of analyzed 
series. 
Possible applications of the introduced 
Multiscale Trend Analysis (MTA) go far 
beyond the linear interpolation problem:
This paper develops and illustrates methods of
self-affine, hierarchical, and correlation 
analyses of time series.
\end{abstract}


\section{Introduction}

~~~~~The motivation for the Multiscale Trend 
Analysis (MTA) introduced in this paper 
is to describe and analyze time series 
in terms of their observed trends
(local linear approximations).
Indeed, trends are the most intuitive 
feature of a time series and it seems 
natural to use them for series 
quantitative description.
Such a description is intrinsically
multiscale since each non-trivial 
process exhibits juxtaposition of 
trends of different duration and 
steepness depending on the observational
scale.

The proposed analysis is based on piecewise
linear approximations of the analyzed time 
series.
Construction of such approximations involves 
a tradeoff between quality and detail.
We formulate (see Sect. \ref{optim}) a local
optimality criterion and use it in a 
multiscale fashion to detect local trends 
in a time series at {\it all} possible scales,
thus forming a {\it hierarchy of trends}.
This hierarchy serves as a unique representation
of the original time series and is used for 
quantitative analysis. 

The problem of piecewise interpolation 
of time series has been given significant 
attention in the context of image processing 
(see for example \cite{SG80,Rob85,Nat91}).
However, the focus was on constructing an optimal 
piecewise linear approximation $L_{\epsilon}(t)$ 
with minimal number of segments for given 
error $\epsilon$ (deviation from the original 
signal).
On the contrary, we concentrate on finding
a whole hierarchy of consecutively more 
detailed approximations.
    
This paper illustrates the following 
applications of MTA:

\begin{itemize}
\item {\it Descriptive and exploratory data 
analysis.} Computationally effective 
trend decomposition naturally complements
a standard data miner's toolbox.
Conveniently, MTA does not rely on any 
assumptions about the analyzed time series 
(e.g. stationarity or existence of 
higher moments) while its results are
easily interpreted 

\item {\it Self-affine analysis.}
Particularly, MTA provides a way to 
extract local fractal properties of 
the processes. 

\item {\it Hierarchical analysis.}
Representation of a time series as a
hierarchy (tree) allows one to use methods 
borrowed from the theory of hierarchical 
scaling complexities \cite{AP}. 
Particularly, Horton-Strahler indexing 
provides a natural way to consider
scaling laws for trends.

\item {\it Correlation analysis.} 
MTA allows one to detect non-linear 
correlations, particularly those caused by 
the presence of amplitude modulation and 
non-linear long-term trends.
\end{itemize}

The paper is organized as follows:
Section \ref{MTD} introduces the basic
notions and describes the computational
algorithm for decomposition of a series
into a hierarchy of trends.
Methods of MTA-based self-affine analysis 
comprise Sect. \ref{Selfaffine}. 
Section \ref{Hier} introduces hierarchical 
analysis of time series.
Correlation analysis is described in Sect.
\ref{Corr}.
Fractional Brownian walks and Mandelbrot
cascade measures are used to illustrate 
methods of Sect. \ref{Selfaffine} - \ref{Corr}.
Section \ref{Discussion} concludes.

\section{Multiscale Trend Decomposition}
\label{MTD}

~~~~~The core of the MTA is construction 
of a hierarchical tree $T_X$ that describes 
the trend structure of a given time 
series $X(t)$ . 
Trend is defined here as a linear least 
square approximation of $X(t)$ at a subinterval
of the observational time interval.
The tree $T_X$ is formed step-by-step, 
from the largest to the smallest scales: 
First, we determine the longer trends, then look for 
the shorter and shorter trends against the background 
of already established ones, all the way down 
the hierarchy of scales.
The larger the scale at which the trend is observed, 
the higher the level of the corresponding vertex 
within the tree.
The root (top vertex) of the resulting tree $T_X$ 
corresponds to the global linear trend of $X(t)$;
each internal vertex corresponds to a distinct 
local trend,
the leaves (vertices with no descendants) to
the the elementary linear segments of the original
time series $X(t)$: $\left[X(t_i),X(t_{i+1})\right]$.
The union of leaves thus coincides with $X(t)$.

A recursive procedure for constructing the 
tree $T_X$ is described below.

\subsection{Scheme of the decomposition}

~~~~~Without loss of generality we presume that the 
time series $X(t)$ is observed at a finite number 
of epochs within the time interval $[0,~1]$. 
At the first step the whole time series 
$X(t),~t\in[0,~1]$ is approximated by a single 
trend --- the linear least square fit $L_0(t)$
(Fig. 1a).

This trend forms the vertex $v^0$ at the level 
0 (the root) of the resulting hierarchical tree $T_X$
(Fig. 1c).
It is also convenient to say that the root of $T_X$ 
corresponds to the whole time interval $[0,~1]$,
and vice versa. 
At the next step we determine secondary trends on the
background of the first global one. 
For this we consider the deviation 
$X_1(t)=X(t)-L_0(t),~t\in[0,~1]$ of $X(t)$ 
from its linear trend $L_0(t)$ and approximate it by a 
piecewise linear (discontinuous) function $L_1(t)$
(Fig. 1b).
The most delicate part of the analysis ---
choosing the optimal number $n^0$ of segments 
for this approximation --- is described below
in Sect. \ref{hystproc}.
The approximation $L_1(t)$ results in partition of the
time interval $[0,~1]=I^0$ into $n^0$ 
nonoverlapping subintervals
$I^1_i=[t^1_i,~t^1_{i+1}],~i=1,\dots,n^0$, with
$t^1_1=0$, $t^1_{n^0+1}=1$.
The linear segments $l^1_i(t)$ that comprise $L_1(t)$ 
are determined by the least square fit 
of $X(t)$ within corresponding subintervals.
They form $n^0$ vertices $v_i^1,~i=1,\dots,n^0$ at 
level 1 of the tree $T_X$. 
The enclosures $I^1_i\subset I^0$ are 
reflected in the structure of the tree $T_X$ by the 
fact that the vertices corresponding to
subintervals $I^1_i$ are descendants of the root, 
which corresponds to $I^0$.  

Repeating the above procedure at arbitrary interval
$I_i^1$ from level 1 we form $n^1_i$ ternary 
linear trends,
each determined by the least square fit of $X(t)$
at a subinterval $I^2_j\subset I^1_i$, $j=1,\dots,n^1_i$.
The union of $N_2=\sum_{i=1}^{n^0} n_i^1$ such trends
descending from all the trends of level 1
form level 2 of the tree $T_X$.
To index the vertices (local trends) 
at level 2 we use the natural ordering 
induced by the corresponding time partition:
$v^2_i$ ($l^2_i$) denotes the vertex (trend) 
that corresponds to the time subinterval 
$I^2_i=\left[t^2_i,~t^2_{i+1}\right]$,
$i=1,\dots,N_2$.

Repeating the same procedure at each time interval 
of level $l$, $l\ge 0$ we form level $(l+1)$.
It consists of 
\[N_l=\sum_{i=1}^{N_{l-1}}n^{l-1}_i\]
subintervals (vertices).
By construction, $N_0=1$ and $N_k<N_p$ for $k<p$. 
We depth of the resulting tree is denoted by $L$. 

Each level $l$ of the tree $T_X$ corresponds 
to a piecewise linear approximation $L_l(t)$ 
of the time series $X(t)$ as well as to the
induced partition 
$I^l=\left\{I^l_i,~i=1,\dots,N_l\right\}$ 
of the observational interval $I^0$.
The global piecewise linear approximation
$L_l(t)$ at level $l$ is a union of local linear 
approximations 
$l^l_i(t)$, $t\in I^l_i=\left[t^l_i,~t^l_{i+1}\right]$,
$i=1,\dots,N_l$, and 
$I^0=\cup_{i=1}^{N_l} I^l_i~\forall l$.

By $r^l_i$ we denote the length of subinterval
$I^l_i$, and by $e^l_i$ the rms deviation
of $X(t)$ from its linear fit $l^l_i(t)$ at this
subinterval:
\begin{equation}
\label{e}
e^l_i=\sqrt{\sum_{t\in I^l_i} \left(\vphantom{^I}
X(t)-l^l_i(t)\right)^2}.
\end{equation}

The total fitting error $E_l$ at the level $l$
is given by
\begin{equation}
\label{E}
E_l^2=\sum_{i=1}^{N_l}\left(e^l_i\right)^2=
\sum_{t\in I^0}\left(
\vphantom{^I}X(t)-L_l(t)\right)^2.
\end{equation}

All vertices (subintervals) at a given 
level of $T_X$ result from the same number of 
divisions of the initial interval $[0,~1]$.
However, in many applications it is desirable
to work with approximations characterized by a
similar scale of observed trends, independently
of the division history.
To take this into account we consider the modified 
tree $M_X$ obtained from $T_X$ by the following 
procedure.
The first two levels of $M_X$ are the same as 
that of $T_X$.
Each consecutive level is formed by division
of only one of the existing subtrends and
leaving all the other unchanged.
A subtrend $v_i^l$ to be divided
corresponds to the maximal improvement
of the fitting quality 
$\Delta=\left(e_i^l\right)^2-\sum \left(e_c^{l+1}\right)^2$, 
where $c$ runs over the indexes of children 
of the vertex $i$.
We will call $T_X$ the topological and $M_X$
the metric tree associated with the series $X(t)$.
To avoid excessive notations we will use the same 
indexing for both the trees $T_X$ and $M_X$ stating 
each time which one is considered.

\subsection{Optimal piecewise linear approximation}
\label{hystproc}

~~~~~Here we describe a procedure for finding 
the optimal piecewise linear approximation 
$L(t)$ of a series $X(t)$ at a given time interval.
Without loss of generality we suppose that the
interval is $[0,~1]$.
The problem, of course, is in finding the optimal 
tradeoff between the number $N$ of linear segments 
within $L(t)$ and the corresponding fitting quality $E$.
Clearly, the larger the number $N$, the better the 
resulting fit. 
Our goal is to depict by linear segments only the most 
prominent large-scale trends of $X(t)$ leaving the smaller
fluctuations for the later steps of the decomposition.
To solve this problem we employ the function
\begin{equation}
\label{hystm}
H(N,E)=-\frac{\log(E/E_0)}{N-1},
\end{equation}
where $E_0$ is the fitting error of the global linear
approximation $L_0(t)$ of $X(t)$ on $[0,~1]$.
This function measures the quality of a piecewise
linear approximation $L(t;N,E)$ which consists
of $N$ linear segments and has total fitting 
error $E$.
The optimal approximation $L(t;N^*,E^*)$ 
corresponds to the maximum of $H(N,E)$:
\begin{equation}
\label{criterion}
H(N^*,E^*)=\max_{N,E}H(N,E).
\end{equation}

Geometrically, consider the plane 
$\left(N,\log(E/E_0)\right)$,
$N$ being the number of linear segments
within a piecewise linear approximation
of $X(t)$ on $[0,~1]$, and $E$ the total 
fitting error.
The global linear approximation $L_0(t)$ 
at the whole interval $[0,~1]$ corresponds to 
the point $p_0=(1,0)$.
An arbitrary piecewise approximation $L_i(t)$ 
corresponds to the point $p_i=(N_i,\log(E_i/E_0))$,
$N_i\ge 1$, $E_i\le E_0$.
The slope of the linear segment $[p_0,p_i]$ shows 
the increase of the fitting quality per one 
additional segment of approximation.
By the criterion (\ref{hystm},\ref{criterion}) 
we chose the approximation with the maximal 
quality increase.  

With the above criterion (\ref{hystm},\ref{criterion}) 
one can find the optimal approximation by a full 
search over all possible partitions of $[0,~1]$ 
by epochs of $X(t)$ into $N=2,3,...$ subintervals.
However, the computational complexity of such an 
approach depends exponentially on the number of 
observations so it can hardly be used in practice.  
In Sect. \ref{optim} below we introduce an optimized 
search based on the idea that partition epochs should 
correspond to the prominent edges of 
the analyzed series $X(t)$.

\subsection{Optimized search}
\label{optim}

~~~~~The idea of the optimized search is to reasonably
reduce the set of possible partition epochs by considering
only those at which $X(t)$ significantly changes 
its slope --- edge points.
 
The edge points are determined by the following 
recursive procedure illustrated in Fig. 2.

At the first step we choose the epochs $(t_1,t_2)$ 
corresponding to the maximum and minimum of the 
detrended function 
$X_1(t)=X(t)-L_0(t)$, 
where $L_0(t)$ is a least-square linear 
fit of $X(t)$ in $[0,~1]$.
If one of these epochs coincides with the
interval boundary (say, $t_1=0$) only the
remaining epoch ($t_2$) is considered.
If both these epochs coincide with the
interval boundaries, we redefine $L_0(t)$ as the 
line connecting $X(0)$ and $X(1)$ and repeat the 
procedure.
As a result we have one or two partition epochs
within the initial interval; they divide it into
two or three subintervals respectively.
The procedure is now repeated for each of these 
subintervals, producing two to six new partition
epochs.
Together with already selected ones, they divide 
the initial interval into, respectively, four 
to nine subintervals, etc.
The partition stops when the predefined number 
$(N_h-1)$ of partition epochs is collected; 
this corresponds to $N_h$ subintervals.

With $(N_h-1)$ possible partition epochs there are 
$\left(2^{N_h-1}-1\right)$ ways to divide the 
interval into $2,\dots,N_h$ subintervals.
The optimal --- according to 
(\ref{hystm},\ref{criterion}) --- partition
can be found by $\left(2^{N_h-1}-2\right)$ operations.

To further reduce the computation volume, 
we first choose the optimal one from 
$(N_h-1)$ partitions formed by $(N_h-2)$ partition 
epochs.
Next, only the $(N_h-2)$ epochs that form this 
partition are used to find the optimal 
partition with $(N_h-3)$ partition epochs, etc.
Finally, we use criterion 
(\ref{hystm},\ref{criterion}) to 
choose the optimal from $(N_h-1)$ partitions, 
each having a distinct number of subintervals 
ranging from $2$ to $N_h$.
This way we reduce the number of operations to 
$(N_h^2-N_h-2)/2$.

Clearly, the above optimization may produce a 
piecewise function which does not 
coincide with the optimal one resulting from 
applying the criterion (\ref{hystm},\ref{criterion}) 
to the whole variety of possible partitions. 
As such, this optimization should be considered as a 
computationally effective approximation of the result.
Extensive numerical experiments show that it is 
reasonably good for a wide range of time series 
including fractional Brownian motions with different 
Hausdorff measures and self-affine processes coming 
from geophysical observations. 

\subsection{Examples}

~~~~~Here we show some examples and illustrate different 
ways to visualize the results of the decomposition.

Figure 3 shows four levels, $l=0,1,2$, and $10$ of 
tree $M_X$ for a fractional Brownian walk with
Hausdorff measure $Ha=0.7$.
Panel a) shows the analysed series $X(t)$
and the piecewise linear approximations
$L_l(t),~l=0,1,2,10$,
while panel b) shows the four corresponding 
levels of the tree $M_X$.

One can see how the fitting quality improves
with the number of linear segments: each
consecutive approximation tries to account
for the most prominent variations of $X(t)$
adding the least possible number of new segments.
For example, starting with the three segments of the 
decomposition $L_1(t)$ at level 1, it is clearly more 
efficient to improve the leftmost segment, 
which exhibits large deviations around $t=0.1$, 
than work with the central or rightmost one.
When work is done with the largest deviation
(see level 2) we proceed to the smaller ones. 

The function shown in Fig. 4a on the background
of its tree $M_X$ is a sum of three
sinusoids with different frequencies.

The amplitudes are chosen in such a way that
the largest fluctuations are carried at the
smallest frequency, intermediate at the 
second largest, and smallest at the highest 
one.
This structure is clearly depicted by
the decomposition with each separate level
responsible for a distinct frequency
(see panels b), c), and d)).

Two more examples are given in Fig. 5 where we 
show only the signals $X(t)$ and the upper 
levels of their trees $M_X$, which is enough
to understand the shape of corresponding
piecewise linear approximations.
Decomposition for the famous Devil's 
Staircase is shown in Fig. 5a:
it gives the exact description of the 
staircase structure. 
Figure 5b shows a decomposition for 
modulated oscillations with time-dependent
frequency.
Contrary to the panel a) here we
use color-code to depict slope changes
(from downward to upward or vice versa), 
not their directions.
In this example one can see how the 
amplitude of oscillation is reflected in the 
decomposition:
the higher the amplitude, the higher
the level at which it is first detected.

\subsection{On the numerical parameter $N_h$}
~~~~~The only numerical parameter of our algorithm 
is the maximal number $N_h$ of secondary trends 
(see Sect. \ref{optim}).
Large values of $N_h$ contribute to the
computational complexity, while small values may
prevent fast detection of optimal approximation
and create superfluous levels of the hierarchy 
$T_X$.
Numerous experiments suggest 
the value $N_h=5$ as the optimal tradeoff, and
we use it for all experiments presented in this
paper.

Clearly, with $N_h=5$ we are not insured from
creating unnecessary levels.
For example the division of Fig. 4b 
consists of 6 ($>N_h=5$) linear segments, 
so it could not be obtained by a single 
division of the original series.
In fact this is level 2 of the original
hierarchy $M_X$.
Analogously, the intermediate division of Fig. 4a
(see also Fig. 4c) corresponds to level 19, and the 
bottom one (Fig. 4d) to level 82.

The simple procedure used to remove unnecessary levels 
is illustrated in Fig. 6 where we show the fitting error 
$E_l/E_0$ for all levels $l$ of the tree $M_X$ 
constructed for the signal of Fig. 4a.
The prominent edge points show the three levels at 
which saturation of the fitting quality is reached;
only these three levels are left in Fig. 4a.

If the analyzed tree has only less-than-5-fold partitions
(which is the case for the Devil's Staircase of Fig. 5a) 
the above procedure is unnecessary.
The properties of this procedure and conditions
for its use are beyond the scope of the present 
paper.

\section{Self-affine analysis}
\label{Selfaffine}

~~~~~In this section we demonstrate how 
self-affine properties of a time series are
reflected in its decomposition $M_X$.

Recall \cite{Mand01,Tur97} that statistical properties of 
a self-affine time series $X(t)$ remain the same 
under the transformation 
\begin{equation}
\label{def}
\left\{\begin{array}{lcl}
t'&=&rt,\\
X'&=&r^{Ha}X.
\end{array}\right.
\end{equation}
That is, when one changes the observational time scale
by a factor of $r$, the scale of measurements should 
be changed by a factor of $r^{Ha}$ in order to preserve
the characteristic statistical features of $X(t)$.
The parameter $Ha$ is called Hausdorff measure;
it is related to the fractal dimension $D$ of a 
self-affine time series as $Ha=2-D$ \cite{Tur97}.
Accordingly, for one-dimensional time series
the Hausdorff measure may take values within 
the range $0<Ha<1$. 
A useful interpretation of $Ha$ comes from the 
character of correlations between the time series
increments: $\Delta_i=X(t_i)-X(t_{i-1})$.
Negative correlations between $\Delta_i$ and 
$\Delta_{i+1}$ lead to high fluctuations of $X(t)$
and as a result to absence of pronounced trends; 
this situation corresponds to  small values of 
Hausdorff measure: $Ha<1/2$.
Positive correlations --- leading to existence of 
long-term trends --- correspond to $Ha>1/2$.
For a process with independent increments
(e.g. Brownian walk) one has $Ha=1/2$.  

To estimate the Hausdorff measure of observed
time series one typically considers the dependence 
of a convenient measure of its variation on 
the length of a corresponding observational 
interval \cite{Mand01,Tur97}.
In our case the appropriate variation measure
can be chosen as the fitting error $E_l$ (\ref{E}) 
of $M_X$ at level $l$.    
According to (\ref{def}), for a self-affine series 
$X(t)$ we expect to observe a power-law relation 
\begin{equation}
\label{EN}
\frac{E_l}{E_0}=N_l^{-Ha}=R_l^{Ha},
\end{equation}
where $N_l$ is the number of segments at the level
$l$, $R_l=N_l^{-1}$ is their mean length. 

As a model example we consider fractional Brownian 
walks (FBWs) with Hausdorff measures in the range 
$0<Ha<1$.

Figure 7a shows trajectories and the corresponding 
$(N_l,E_l)$-scalings for three FBWs with 
$Ha=0.1, 0.5,$ and $0.9$. 
Figure 7b shows the value $b(Ha)$ estimated 
by the best linear square fit from the relation 
\begin{equation}
\label{data}
\log(E_l/E_0)=-b\log(N_l)
\end{equation}
based on decomposition of 2100 independent FBWs;
to remove statistical fluctuations we averaged
$b$ over 100 FBWs for each value of $Ha$.
As seen in Fig. 7b, the scaling (\ref{EN}) 
clearly holds for $Ha>0.3$; 
the deviations observed at the smaller values 
of $Ha$ are due to the fact that the corresponding
FBWs become noisier and hardly display pronounced 
trends.
This effect is typical for self-affine analysis
(e.g., see \cite{DFA}).
To neglect it we consider the integrated signal 
$Y(t)=\sum_{s\le t} X(s)$.
Estimations of the slope $b(Ha)$ 
for integrated FBWs are presented in Fig. 7c.
The linear relation $b(Ha)=Ha+1$ is now observed 
for $0<Ha<0.6$, the change of slope
compared to Fig. 7b is due to the integration 
procedure.

Another way to estimate $Ha$ is to 
consider the error-length dependence for all 
individual linear segments comprising $M_X$:
\begin{equation}
\label{el}
\frac{e^l_i}{E_0}=\left(r^l_i\right)^{Ha+1/2},
~l=1,\dots,L,
~i=1,\dots,N_l.
\end{equation}
The difference in the power exponents of
relations (\ref{EN}) and (\ref{el}) is explained 
by the fact that the former deals with averaged 
statistics, while the latter deals with 
characteristics of individual intervals.
Figure 8 illustrates the error-length 
dependence (\ref{el}) for FBWs with
$Ha=0.1$ and $Ha=0.9$.

Importantly, MTA provides a convenient basis 
for estimation of local Hausdorff measures $Ha(t)$.
Consider all the intervals from $T_X$ that 
cover epoch $t$. 
At each level $l$ of $T_X$ there is one and 
only one such interval; we use the index $^l_{(t)}$
to denote this interval and all its characteristics.
The local Hausdorff measure $Ha(t)$ is
estimated now from the relation
\begin{equation}
\label{Haloc}
\frac{e^l_{(t)}}{E_0}=\left(r^l_{(t)}\right)^{Ha(t)+1/2},
~l=1,\dots,L.
\end{equation}
Figures 9a,b show the dynamics of the local
Hausdorff measure for multi- and monofractals.
We use a Mandelbrot cascade measure 
$M(0.7,0.3;0.3,0.7)$ as a model example of a
multifractal (Fig. 9c), and a Brownian walk 
as that of a monofractal (Fig. 9d). 
The definition of Mandelbrot cascade measure
is given in Appendix \ref{mcm}.
Note that the range of $Ha(t)$ variation 
for the monofractal (Fig. 9b) is an order 
of magnitude less than that for the 
multifractal (Fig. 9a).

The points $\left(e^l_{(t)},~r^l_{(t)}\right)$ 
used in (\ref{Haloc}) to estimate the local
Hausdorff measure are extracted from
the whole set $\left(e^l_i,~r^l_i\right)$ of
(\ref{el}).
This suggests a method for detecting multifractality
in $X(t)$: the larger the scattering of the
points $\left(e^l_i,~r^l_i\right)$, the larger
the probability that the observed series
is a multifractal.
Formal statistical tests can be easily constructed
from this general principle based on the particular
problem at hand.
The character of temporal variations of $Ha(t)$ 
(Figs. 9a,b) can be also used in such tests.
An example of the scattering 
$\left(e^l_i,~r^l_i\right)$ is shown in 
Fig. 10 for mono- and multifractals of Fig. 9.
In this model example the difference is obvious.

\section{Hierarchical scaling}
\label{Hier}

~~~~~The appropriate ordering of vertices 
within a tree $T_X$ is very important for 
meaningful description and analysis 
of the series $X(t)$. 
The problem of such an ordering becomes
not trivial as soon as the tree is not 
uniform (i.e. is not formed by
applying the same deterministic division 
rule to each of its vertices).
A befitting way to solve this problem is
given by the Horton-Strahler topological 
classification of ramified patterns
\cite{AP,Horton,Strahler} 
illustrated in Fig. 11:
One assigns orders to the vertices of the tree, 
starting from $k=1$ at leaves (vertices with no 
descendants).

The order of an internal vertex equals 
the maximal order $m$ of its descendants, 
if they are distinct, and $m+1$ if they are 
all equal.
Originally introduced in geomorphology 
by Horton \cite{Horton} and later refined by 
Strahler \cite{Strahler}, this classification 
is shown to be inherent in various geophysical, 
biological, and computational applications 
\cite{AP,NTG97,GNT99,Toro01}.  
 
As a result of the Horton-Strahler indexing of
the tree $T_X$, 
each of its vertices is characterized by an order $k$, 
length $r$ of the corresponding partition interval, 
and the error $e$ of the linear least square fit of 
$X(t)$ on this interval. 
The scaling behavior of $X(t)$ can be described by 
the exponents of the relations:
\begin{equation}
\label{HSscaling}
N(k)\sim 10^{-B_N k};
~R(k)\sim 10^{B_R k};
~E(k)\sim 10^{B_E k}.
\end{equation}
Here $N(k)$ is the number of vertices of order $k$,
$R(k)$ and $E(k)$ are the values of $r$ and $e$ 
averaged over the vertices of order $k$. 

The relation between the number $N(k)$ of 
vertices of order $k$ and their average length 
$R(k)$ determines the fractal dimension $d$ 
of the tree $T_X$ \cite{NTG97}:
\begin{equation}
\label{fdim}
N(k)=R(k)^{-d}.
\end{equation}
Combining (\ref{HSscaling}) and (\ref{fdim})
we find:
\begin{equation}
\label{dimrel}
d=\frac{B_N}{B_R}.
\end{equation}

The structure of the tree $T_X$ can be considered
at different levels of detail: 
First, one can consider only the topological 
structure (Fig. 12a),
where the position of each vertex is uniquely 
determined
by its parent (the nearest vertex placed closer to
the root); and any permutation of siblings
(the vertices with the same parent) does not change
the tree.
Each vertex is characterized by its 
Horton-Strahler index, 
and the only constraint on a tree 
resulting from MTA is the maximal possible number 
$N_h$ of siblings, that is subtrends within a
given trend.
Next, one can add the information on interval 
partition (Fig. 12b): 
The siblings become ordered according to the 
partition of the interval corresponding 
to their parent. 
Each vertex $v_i$ is additionally 
characterized by the length $r_i$ and the 
following conservation law holds:
\begin{equation}
\label{lcons}
r_i=\sum r_c,
\end{equation}
where $c$ runs over the indexes of the 
children of the element $i$.

Finally, (Fig. 12c) one considers error 
characteristics $e_i$, 
which describe the quality of the linear fit 
of $X(t)$ within the corresponding time interval.
In terms of these errors the system becomes
dissipative:
\begin{equation}
\label{diss}
e_i\ge\sum e_c,
\end{equation}
with the same meaning of subindexes as in 
(\ref{lcons}).

The exponents $B_{N,R,E}$ of (\ref{HSscaling})
reflect different statistical properties of the
tree $T_X$: 
$B_N$ describes its topological structure while
$B_R$ and $B_E$ relate to the metric structures 
based, respectively, on properties of interval 
partition ($r$-metric) and piecewise linear fit
($e$-metric).

For illustration we again use FBWs with 
different Hausdorff measures.

Figure 13 shows the dependence of the exponents 
$B_{N,L,E}$ on the Hausdorff measure 
$0\le Ha \le 1$.
The estimations are averaged over 100 FBWs for
each value of $Ha$.
The exponents $B_N$ and $B_R$ are nearly constant:
$B_N\approx 0.52$, $B_R\approx 0.57$, while
for the exponent $B_E$ we observe the linear
dependence:
\begin{equation}
\label{EHa}
B_E=0.7+Ha\approx \log_{10}(5)+Ha.
\end{equation}

These results have an important interpretation:
All FBWs with Hausdorff measure in the range
$0\le Ha \le 1$ have the same topological and 
$r$-metric structures in terms of MTA tree $T_X$.
Particularly, trees $T_X$ corresponding to 
different $Ha$ have the same fractal dimension 
$d=B_N/B_R\approx 0.9$. 
The only characteristic that depends on the 
Hausdorff measure is the fitting error 
($e$-metric), that is the degree of variation of 
$X(t)$ within a given interval.

\section{Correlation analysis}
\label{Corr}

~~~~~One of the important applications of MTA 
is correlation analysis of time series.
The major drawback of classical correlation 
analysis is that interpretation of its results 
may be completely ruined by the presence of 
long-term trends and/or modest amplitude 
modulations of signals.
The MTA can naturally avoid these problems
by depicting the essential local properties of 
the analyzed series.

We start this section by introducing two
measures of similarity for time series.
One is based solely on the time interval partition 
induced by $M_X$;
another takes into account the directions 
(upward vs. downward) of local trends.

\subsection{Distance between partitions}
~~~~~Each level $l$ of the tree $M_X$ 
(Sect. \ref{MTD}) corresponds to a partition of 
the time interval $[0,~1]$ into $N_l$ nonoverlapping 
subintervals.
Since each of these subintervals corresponds to a 
distinct observed trend of the series $X(t)$, the 
problem of comparison of two such partitions 
naturally arises.
Below we introduce the distance between two 
partitions.

Consider the space $\Omega$ of finite partitions 
of the unit interval $[0,~1]$.
Each partition $A$ is defined by a finite 
number $n_A$ of points; the boundaries 
$0$ and $1$ are included in all partitions:
\[A=\{0=a_0<a_1<\dots< a_{n_A}<a_{n_A+1}=1\}.\]
The trivial partition $U$ consists only of
boundary points: $U=\{0,~1\}$.

For $A,B \in \Omega$ we say that $B$ is a 
subpartition of $A$ ($B\subset A$) if all
points from $A$ are among points from $B$; 
this imposes a partial order on $\Omega$.
A union $A\cup B$ is defined as the partition 
consisting of the points included in either
$A$ or $B$, without repetitions.
An intersection $A\cap B$ is defined as 
the partition consisting of points included in 
both $A$ and $B$.    

An asymmetric distance $m(A,B)$ from $A$
to $B$ ($A,B\in\Omega$) can be defined as
\begin{equation}
\label{dist1}
m(A,B)=\sum_{i=1}^{n_A}\min_{0\le j\le n_B+1}
\{|a_i-b_j|\},
\end{equation}
which gives for the trivial partition
\[m(A,U)\equiv m(A)=\sum_{i=1}^{n_A}
\min\{a_i,1-a_i\}\]
The distance (\ref{dist1}) is interpreted as 
the minimal correction 
to $A$ that makes $B$ its subpartition:
$B\subset A'$, where $A'$ stands for the
corrected version of A.

The following properties of $m(A,B)$ 
follow directly from the definition 
(\ref{dist1}):
\begin{enumerate}
\item $0\le m(A,B)<\infty$; 
\item $m(A,B)=0$ iff $B\subset A$;
\item Additivity with respect to $A$:
$m(A_1\cup A_2,B)=m(A_1,B)+m(A_2,B)$;
\item Monotonicity with respect to $B$ 
(the triangle inequality):
$m(A,B_1\cup B_2)\le m(A,B_1)+m(A,B_2)$.
\end{enumerate}

It is convenient to consider the 
symmetric function 
\begin{equation}
\label{dist2}
\mu(A,B)=\max\{m(A,B),m(B,A)\},
\end{equation}
whose small values signal that the partitions
$A$ and $B$ are similar. Note that $\mu$ is
not a distance since it does not satisfy 
the triangle inequality.  
The reciprocal $\mu^{-1}$ may serve as a 
measure of partition correlation.

\subsection{Slope sign correlation} 
\label{ssc}
~~~~~Here we introduce the correlation function
that describe similarity between two piecewise 
linear approximations $L^1(t)$ and $L^2(t)$ 
of $X(t),~t\in [0,~1]$.
(We use upper indexes in order not to mix
these arbitrary approximations with $L_1(t)$,
and $L_2(t)$ at the first and second levels of
the decomposition.)   
This correlation function is based on the 
coarse information about trends from $L^i(t)$: 
We take into account only their directions --- 
upward vs. downward.

First, we introduce the signed partitions
$P_1$ and $P_2$ of the interval $[0,~1]$.
They are formed by the intervals of constant
sign of the slope of $L^i(t),~i=1,2$ 
(see Fig. 14a).

A subinterval from $P_i$ is 
assigned the sign "+" if the corresponding 
trend of $L^i(t)$ is upward, and "--" if it 
is downward.  
Second, we define the signed partition $P$ 
as a union of $P_i,~i=1,2$ with the signs 
determined by multiplication of the signs 
of the corresponding subintervals from $P_i$
(Fig. 14b).
As a result, the positive intervals of $P$
correspond to matching (up to direction) trends 
of $L^1$ and $L^2$, while negative to unmatching 
ones.   

Each subinterval $I$ of the partition 
$P$ is formed by intersection of two 
subintervals $I_i\in P_i,~i=1,2$; 
two general variants of such an 
intersection are shown in Fig. 14c.
A subinterval $I$ is assigned a triplet
$(a,b,c)$ defined as shown in Fig. 14c:
$b$ is the length of the intersection 
$I_1\cap I_2$, while $a$ and $c$ are the 
lengths of those parts of $I_i$ that
are not included in the intersection.
The triplet is normalized:
$a+b+c=1$.
It describes how good is the
matching of intervals $I_i$: the meaning
of $b$ is clear; the best matching for
a given $b$ corresponds to the case when 
the intervals' ends coincide, that is to
$a\cdot c=0$.
The matching quality can be reflected in 
the weight
\begin{equation}
\label{weight}
w=-\frac{(1-b)\log(1-b)}{a\log(a)+c\log(c)}=
-\frac{(a+c)\log(a+c)}{a\log(a)+c\log(c)},
\end{equation}
lying within the range $0\le w \le 1$.

The correlation function $r(L^1,L^2)$ is
now defined as
\begin{equation}
\label{corrf}
r(L^1,L^2)=\sum_{k} r_k\cdot w_k.
\end{equation}
Here the summation is taken over all the
subintervals of the signed partition $P$;
$r_k$ denotes the signed length of
the $k$th subinterval, $w_k$ is the 
corresponding weight (\ref{weight}).

The measure (\ref{corrf}) is intentionally 
crude: it does not distinguish between 
steepness of the trends. 
More elaborate correlations can be easily 
defined following the scheme outlined above.
Nevertheless, as we show in 
Sect. \ref{exs} below, even the roughest 
measure (\ref{dist1}) is very effective
in detecting non-linear correlations.

\subsection{Examples}
\label{exs}
~~~~~This section illustrates applications 
of the correlation analysis in the presence 
of long-term nonlinear trends and amplitude 
modulations.

\subsubsection{Detection of correlation}
\label{detect}
~~~~~Figure 15 displays the trajectories of
two processes $F_i(t),~i=1,2$ coupled by the
common underlying phenomenon which --- 
by and large --- makes them change their 
intermediate-scale trends synchronically.
The most striking similarity between
$F_i(t)$ is observed at the intervals 
$[0,0.1]$ and $[0.2,0.55]$.
Also we note the synchronous peaks around
$t=0.675,0.775,0.975$ (more
pronounced for $F_1(t)$.)
At the same time, the coupling phenomenon
is not a primary one in shaping the dynamics 
of $F_i(t)$, so their overall outlooks are 
still quite dissimilar.
In such situations one is interested in 
detection and proper quantification of the 
observed non-linear coupling. 
The problem of such a quantification 
constitutes an important part of modern 
analysis of time series.

MTA suggests an effective way of 
solving this problem by comparing the
trend structures of observed series at
different scales.
We decompose the observations $F_i(t)$
into trees $M_i$ and calculate the
distance $\mu$ (\ref{dist2}) between 
different levels of these decompositions.
The reciprocal $\mu^{-1}$ of the distance
between the signals $F_i(t)$ 
is plotted as the function of the 
decomposition levels
$l_i,~i=1,2$ in Fig. 16a.

The diagonal ridge indicates pairs of 
levels with similar trend structures.
The prominent upwell observed at
the medium scales --- 
$15\le l_1\le 18,~14\le l_2\le 17$ 
--- signals that this range is responsible 
for the observed coupling.
The maximum $\mu^{-1}=4.6$
corresponds to the levels $l_1=15,~l_2=14$;
we will refer to them as levels of
maximal correlation (LMC).
The piecewise linear approximations 
of $F_i(t)$ corresponding to the LMC 
are shown in Fig. 17.
They clearly accentuate the observed 
coupling.

A typical shape of $\mu^{-1}$
for uncoupled time series is shown
for comparison in Fig. 16b.
The diagonal ridge is still observed,
though it is more blurred.
Existence of such a ridge is explained 
by the fact that partitions with a similar
number of segments, even non-matching 
ones, are closer to each other in the sense
of (\ref{dist2}) than partitions with
significantly different number of segments.
Comparing Figs. 16a and b we conclude that
the upwell observed in panel a is not a random
one and is due to the correlation between
the signals.
A formal statistical test for establishing 
the significance of the observed peaks 
of $\mu$ can be easily constructed.

\subsubsection{Quantification of detected correlation}
\label{quant}
~~~~~As was shown in the previous section,
MTA allows one to estimate non-linear correlations 
between signals; the value $\mu^{-1}$ may be
considered as a measure of such correlation.
Here we show how to evaluate the functional
form of the coupling phenomenon responsible
for the correlation detected.

To pose the problem formally, suppose that the 
observations 
$F_i(t),~i=1,2$ are formed by applying 
amplitude modulations $A_i(t)$ and
adding non-linear trends $T_i(t)$ 
to the same base signal $X(t)$:
\begin{equation}
\label{nlt}
F_i(t)=A_i(t)\cdot X(t)+T_i(t)+\xi_i(t),~i=1,2.
\end{equation}
Here $\xi_i(t)$ are measurement errors.
In this model the correlation between
signals $F_i(t)$ is due totally to the $X(t)$.
The first problem is to reconstruct trends $T_i(t)$ 
and modulated signals $A_i(t)\cdot X(t)$ given 
the observations $F_i(t)$.
Clearly, for reliable reconstruction one has to 
assume an appropriate rate of variation for the 
trends as well as a reasonably small noise-to-signal 
ratio.
In practice, we assume that such conditions are 
satisfied if significant 
coupling has been detected by the correlation 
analysis of Sect. \ref{detect}.

The idea of reconstruction is that the 
correlated parts
$A_i\cdot X(t)$ should be described by the 
LMC of $M_i$ (see Sect. \ref{detect}).  
Accordingly, the trends $T_i(t)$ should be 
described by the higher-scale (less detailed) 
levels. 

As a model example we again use the series
of Fig. 15; in fact, they are produced by 
the model (\ref{nlt}) with
\begin{eqnarray}
\label{funcs}
X(t)   & = & \sin\left(400\pi t(t-0.5)(t-0.7)(t-1)\right); 
\nonumber \\
T_1(t) & = & 5\sin\left(4\pi t^{3/2}\right);\nonumber\\~
T_2(t) & = &-5\cos\left(2\pi t^{3/2}\right);\nonumber\\
A_1(t) & = &\exp(2t);\nonumber\\~
A_2(t) & = &2\exp\left(-t/3\right).
\end{eqnarray}
The measurement errors $\xi_i(t)$ are modeled 
by independent Brownian walks so they 
also represent random drifts.
The series $F_i(t)$ together with their 
components (\ref{funcs}) are shown in 
Fig. 18.

The trends $T_i(t)+\xi_i(t)$ are estimated 
by the piecewise linear functions $\hat T_i$, 
formed by the parents of the vertices 
at the LMC, $l_1=15,~l_2=14$.
In other words, each of the linear 
segments at the levels $l_i$ should be 
formed by a single non-trivial partition 
of one of the trends of $\hat T_i$.
By single we mean that this is a one-time
partition by the rules described
in Sect. \ref{MTD};
by non-trivial --- that each segment 
is divided into more than one 
subsegment. 
The modulated signals $A_i(t)\cdot X(t)$ are 
estimated then as
$\widehat{A_iX_i}(t)=\left(F_i(t)-\hat T_i(t)
\right),~i=1,2.$

The quality of these estimations is illustrated
in Fig. 19 where we show real vs. estimated
modulated signals $A_iX_i(t)$.
The estimations are almost perfect at the
intervals $[0,0.1]$ and $[0.2,0.55]$,
(cf. Fig. 15 and its discussion in Sect. 
\ref{detect}.)
Generally, we catch well the oscillatory
structure of the signals; that is their 
time-dependent frequencies and directions
(upward vs. downward), while the amplitude
estimation is less precise.

The estimations of Fig.19 can be further
improved by means of various kernel smoothing
techniques.
MTA results can be used for optimization 
of the time-dependent kernel width.   

With additional assumptions about the rate
of variation for $A_i(t)$ one may pose the
problem of reconstructing $X(t)$ given
two, or more, modulated versions 
$A_i(t)\cdot X(t)$.
Using the epochs assigned to the 
summands of (\ref{dist1}) (say, $a_i$), 
one may analyze time-dependent correlations 
within $F_i(t)$.
Clearly, the entire analysis can be
repeated with the correlation (\ref{corrf})
as a measure of trend similarity.

\section{Discussion}
\label{Discussion}
~~~~~The methods developed in this paper are
based on the computational technique (see Sect. \ref{MTD}) 
for solving the linear interpolation problem for time series.
This problem includes two principal difficulties.
The first is a fundamental one:
a tradeoff between the quality of a possible 
approximation and its detail.
The second difficulty is purely computational: 
There are $(n-2)!/(n-1-k)!(k+1)!$ ways to construct
a piecewise linear approximation with a given 
number $k$ of segments and $n$ observational epochs.
Clearly, the search for the optimum over all possible 
approximations is unacceptable for operational use, 
and computationally effective algorithms are to
be invented. 
Here we resolve the first difficulty by introducing 
the optimality criterion (\ref{hystm},\ref{criterion}) 
of Sect. \ref{hystproc}, 
and the second by replacing the original time series with
its "skeleton" that includes only the edge points defined
in Sect. \ref{optim}.
The whole analysis is then done hierarchically, 
in a multiscale self-similar fashion. 
This contributes to computational efficiency 
as well as to the imprecision of the final result, 
since the errors made in the first steps of
the decomposition may affect all the consecutive steps.
It would therefore be interesting to study
a) deviations of the MTA approximations 
from the optimal (in a squared deviation sense) 
piecewise linear approximations with the same
number of segments,
and 
b) the history of the first-step errors.

The procedure for edge point detection 
is introduced here (Sect. \ref{optim}) in its
simplest (not to say most naive) form and is 
subject to further improvement.
Nevertheless, even in its present form,
the MTA has the potential to be an effective 
tool for solving a wide specrtrum of applied 
problems, ranging from exploratory data analysis
to studying hierarchical scaling for time series.

Recently, several techniques based on properties of 
local linear trends were proposed and studied.
The Detrended Fluctuation Analysis (DFA) \cite{DFA} 
was shown to be a powerful tool for multiscale
analysis and interpretation of diverse medical
and financial data. 
Contrary to our analysis, DFA uses a predefined 
interval partition scheme independent of the 
particular series at hand.
It is oriented toward analysis of variations, rather 
than the trend structure itself.
An alternative approach to the problem of detection
of local linear trends is discussed in \cite{CS90}.

The problem considered in this paper naturally 
extends to higher dimensions. 
However, it is not clear how to apply the ideas 
developed here even to 2D and this issue deserves 
special attention.
Interestingly, elegant theoretical results on
rectifiable curves by P. Jones \cite{Jones} are 
tightly related to detection of linear structures 
in point clouds.
Various methods of multiscale geometric analysis
based on Jones' theory (\cite{Lerman} and references
therein) use predefined (dyadic) partition schemes.
It would be very important to find algorithms
for fast linearization in point clouds.

It is worth mentioning that the self-affine analysis 
of Sect. \ref{Selfaffine} may be done equally effectively 
by a multitude of techniques, and MTA is by no means
claimed to be the most efficient one. 
We include this section in order to demonstrate the
diversity of possible applications based on the
single MTA decomposition of a time series.
\bigskip

{\bf Acknowledgments.}
We are grateful to Robert Mehlman for valuable
discussion and 
David Shatto for help in preparation of this paper.
This work was supported by a Collaborative Activity 
Award for Studying Complex Systems from the 21st
Century Science Initiative of the James S. McDonnell
Foundation and INTAS 0748.

\newpage
\appendix

\section{Mandelbrot cascade measures}
\label{mcm}
A Mandelbrot cascade measure 
$M(r_i,m_i)$, $i=1,\dots,n$ 
on the interval $[0,~1]$ is constructed 
as follows.
At step 0 there is a unit mass distributed 
uniformly over the whole interval.
At the first step we divide the interval
$[0,~1]$ into $n$ subintervals of lengths
$r_i$, $\sum_{i=1}^n r_i=1$
and assign to them masses $m_i$,
$\sum_{i=1}^n m_i=1$.
Within each interval the mass distribution 
is uniform.
Next, we divide each subinterval $i$
into $n$ subsubintervals and assign
to them uniform masses $m_i\cdot m_j$,
$j=1,\dots,n$, and so on.
Therefore, at the $k$th step the interval
$[0,~1]$ is divided into $n^k$ subintervals,
each carrying the uniform mass
$m_{i_1}\cdot\dots\cdot m_{i_k}$, with $i_k$ taken
from the set $1,\dots,n$ with possible 
repetitions.

Such measures were introduced first to model
turbulent dissipation, and were studied by 
Mandelbrot \cite{Mand74}.

\newpage
\begin{figure}[p]
\centering \includegraphics{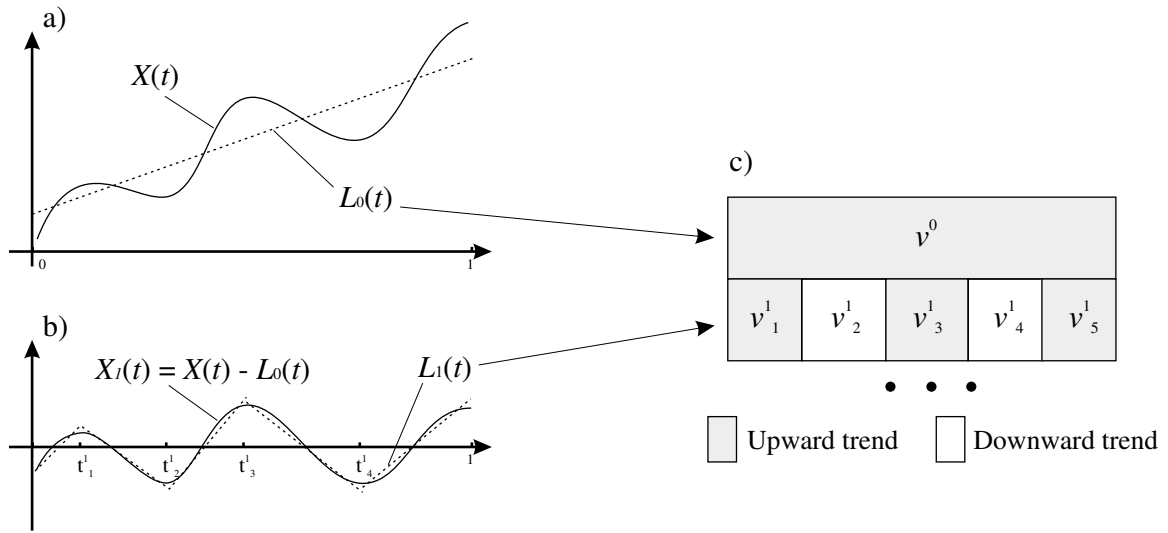}
\caption{
Scheme of the Multiscale Trend Decomposition.
a) At zero step $X(t)$ is approximated by its
global linear trend $L_0(t)$.
b) Detrended series $X_1(t)=X(t)-L_0(t)$ is 
approximated by the piecewise linear 
function $L_1(t)$, 
the whole analysis is then repeated at each of 
subintervals $[t_i^1,~t_{i+1}^1]$.
c) Resulting hierarchy of trends. 
See Sect. \ref{MTD} for details.}
\end{figure}
\clearpage

\newpage
\begin{figure}[p]
\centering \includegraphics{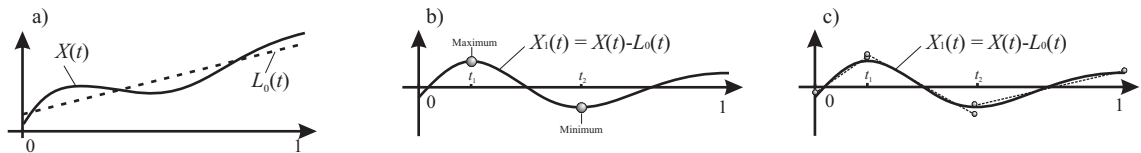}
\caption{Scheme of detection of edge points.
a) At zero step $X(t)$ is approximated by its
global linear trend $L_0(t)$.
b) Epochs $(t_1,t_2)$ of global maximum and minimum
of the detrended series $X_1(t)=X(t)-L_0(t)$ are
located.
c) Analysis is repeated at each of subintervals
$[0,t_1]$, $[t_1,t_2]$, and $[t_2,1]$.}
\end{figure}
\clearpage

\newpage
\begin{figure}[p]
\centering \includegraphics{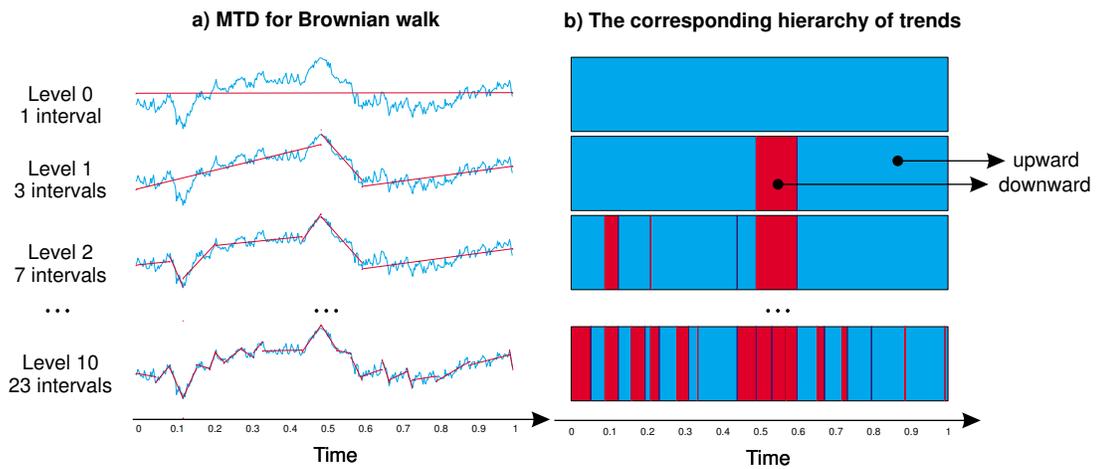}
\caption{Decomposition of a Fractional Brownian walk
with Hausdorff measure $Ha=0.7$.
a) Piecewise linear approximations at levels
$l=0,1,2,10$.
b) Corresponding hierarchical tree.}
\end{figure}
\clearpage

\newpage
\begin{figure}[p]
\centering \includegraphics{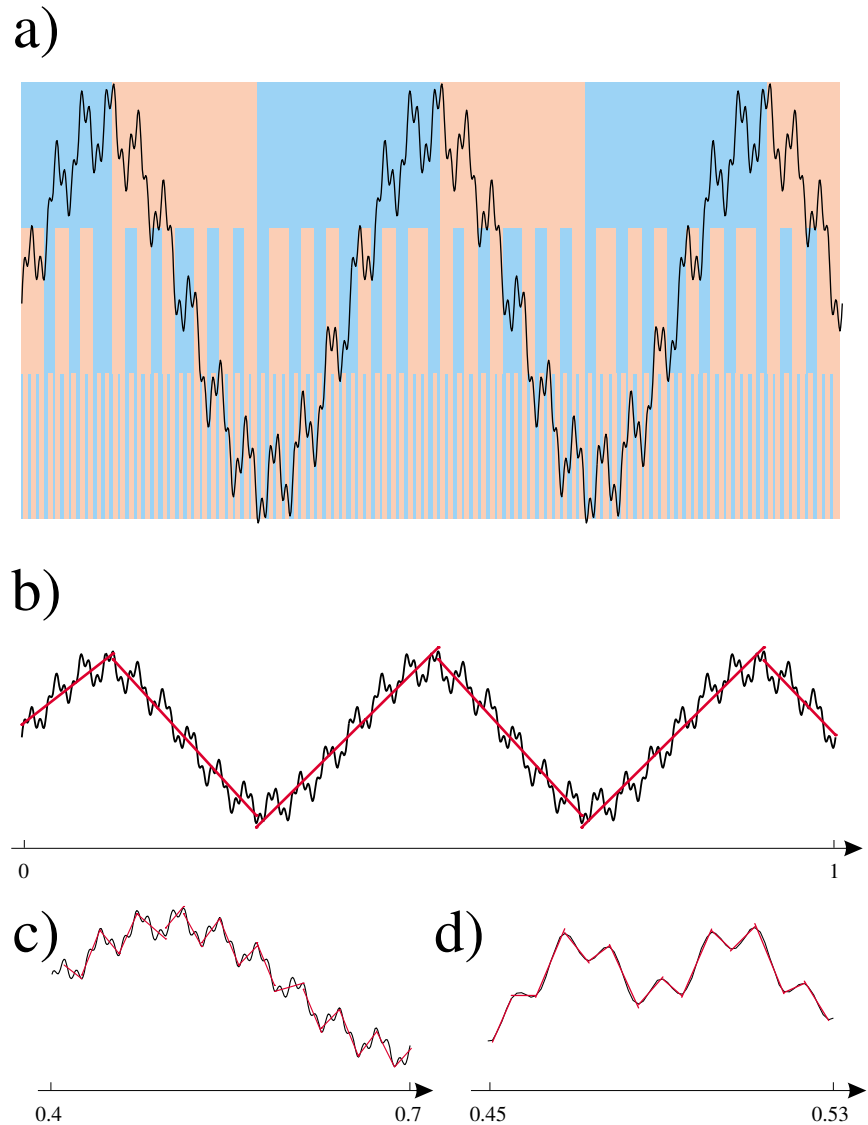}
\caption{Decomposition of the sum of 
three sinusoids, 
$X(t)=\sin(5\pi t)+
\frac{1}{5}\sin(60\pi t)+
\frac{1}{10}\sin(200\pi t)$.
a) $X(t)$ on the background of
three levels from its decomposition.
b) Piecewise linear approximation
corresponding to the top level
of the decomposition shown in panel a).
c) Fragment corresponding to
the middle level of a).
d) Fragment corresponding to
the bottom level of a).}
\end{figure}
\clearpage

\newpage
\begin{figure}[p]
\centering\includegraphics{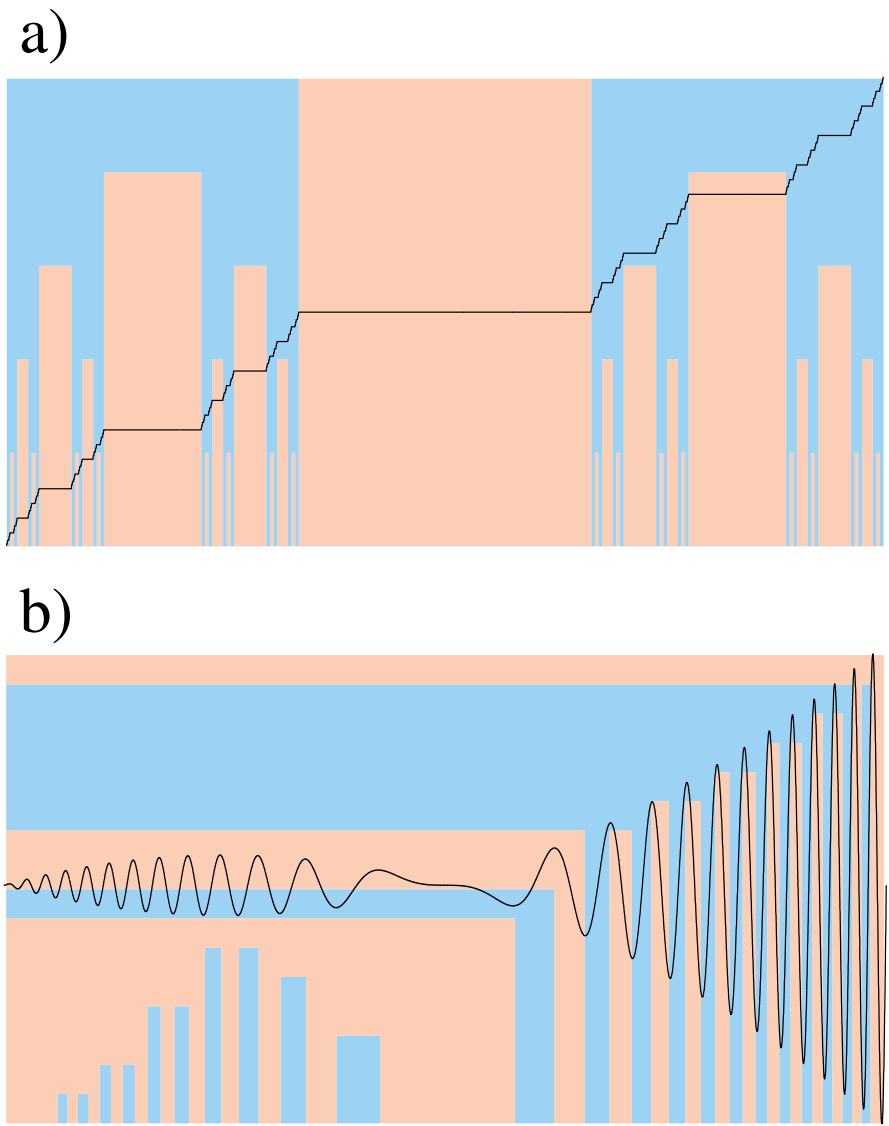}
\caption{Decomposition of
a) Devil's Staircase (5 upper levels 
of $M_X$ are shown) and
b) modulated sinusoid with 
time-dependent frequency (15 levels
are shown).}
\end{figure}
\clearpage

\newpage
\begin{figure}[p]
\centering \includegraphics{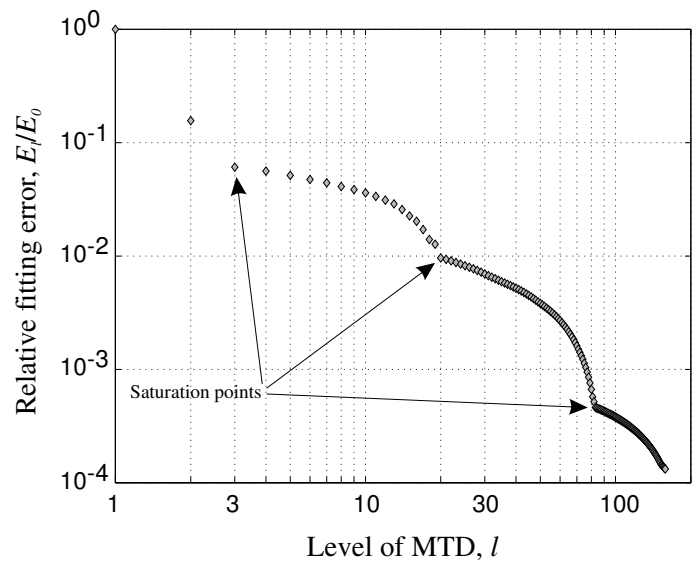}
\caption{Illustration of removing unnecessary 
levels from the decomposition (for the signal 
shown in Fig. 4).
Prominent saturation points 
correspond to the three levels 
shown in Fig. 4}
\end{figure}
\clearpage

\newpage
\begin{figure}[p]
\centering \includegraphics{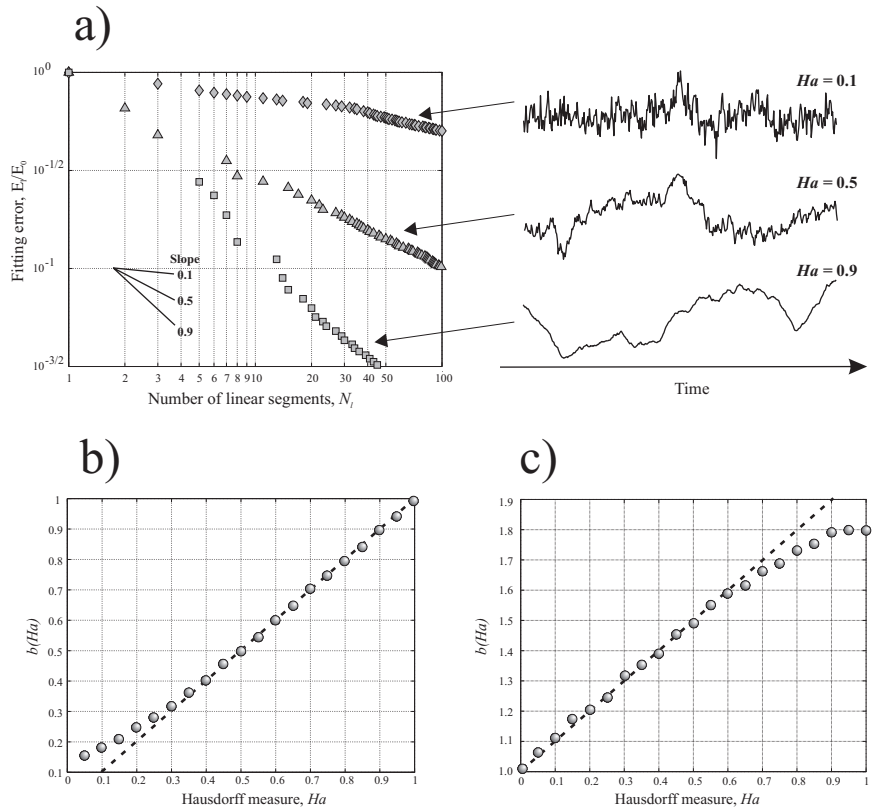}
\caption{Relation between Hausdorff measure 
and error scaling for fractional Brownian 
walks (FBW).
a) Trajectories of FBWs with 
$Ha=0.1,0.5,0.9$ and corresponding
error scalings.
b) Relation $b(Ha)$ for FBWs,
$0\le Ha\le 1$, values of $b$ 
averaged over 100 realizations of FBW
for each value of $Ha$.
c) The same as b) for integrated FBWs.}
\end{figure}
\clearpage

\newpage
\begin{figure}[p]
\centering \includegraphics{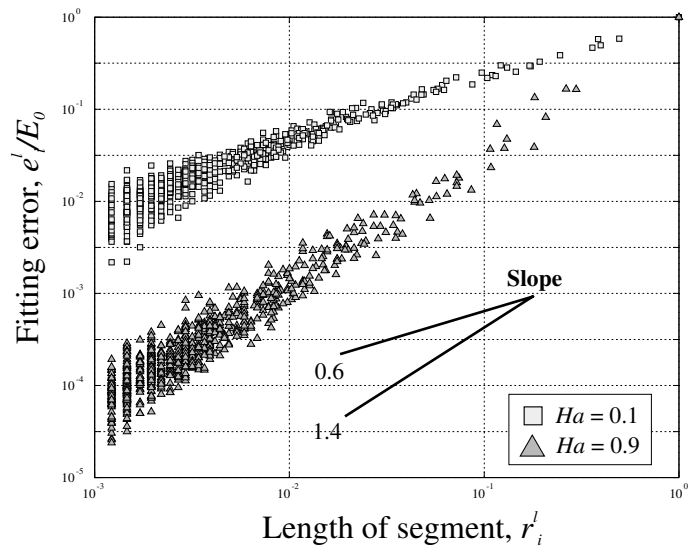}
\caption{Error-length dependence for 
individual vertices of trees $M_X$
corresponding to FBW with $Ha=0.1,0.9$.
The scaling (\ref{el}) is clearly 
observed. }
\end{figure}
\clearpage

\newpage
\begin{figure}[p]
\centering\includegraphics{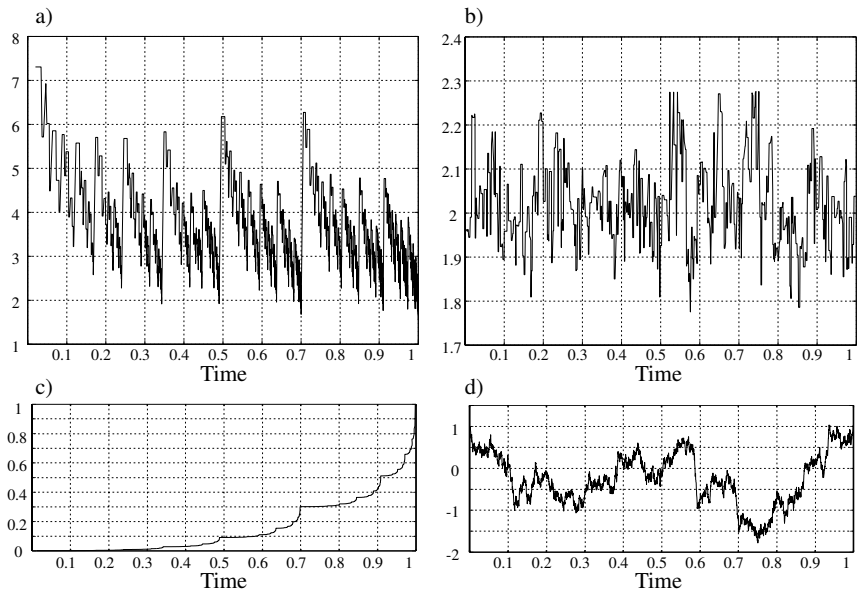}
\caption{Estimation of local Hausdorff
measures, $Ha(t)$ for a multifractal 
(Mandelbrot cascade measure) (panel a)
and monofractal (Brownian walk) 
(panel b).
Corresponding time series are shown
in panels c) (multifractal) and d) 
(monofractal).}
\end{figure}
\clearpage

\newpage
\begin{figure}[p]
\centering\includegraphics{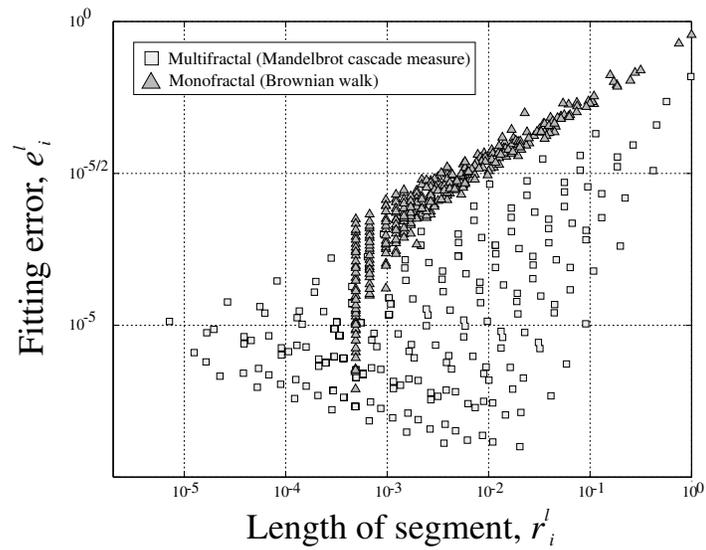}
\caption{Error-length dependence for multi-
and monofractals of Fig. 9.
Note that the point scattering is
significantly larger for the 
multifractal.}
\end{figure}
\clearpage

\newpage
\begin{figure}[p]
\centering \includegraphics{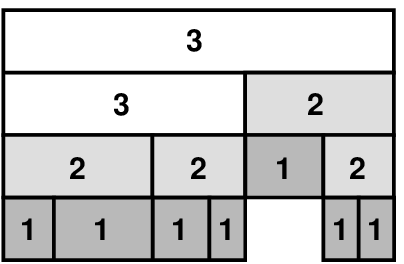}
\caption{Horton-Strahler indexing.}
\end{figure}
\clearpage

\newpage
\begin{figure}[p]
\centering \includegraphics{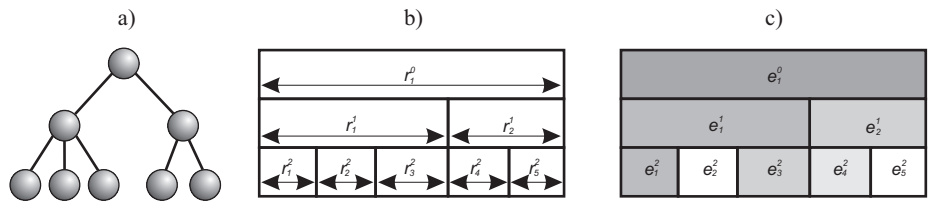}
\caption{Three levels of detail in
MTA description of a time series.
a) Topological.
b) r-metric, based on the interval 
partition.
c) e-metric, based on local linear 
fit of the series.
See details in Sect. \ref{Hier}.}
\end{figure}
\clearpage

\newpage
\begin{figure}[p]
\centering \includegraphics{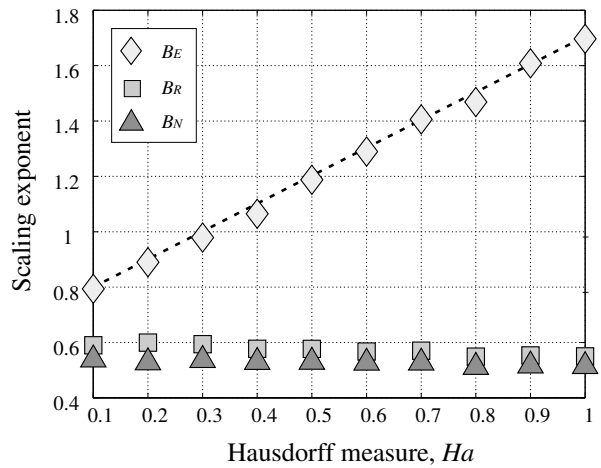}
\caption{Dependence of scaling exponents 
$B_{E,R,N}$ (Eq. (\ref{HSscaling}))
on the Hausdorff measure $Ha$ of
FBWs.
Dashed line is $B=0.7+Ha$.}
\end{figure}
\clearpage

\newpage
\begin{figure}[p]
\centering \includegraphics{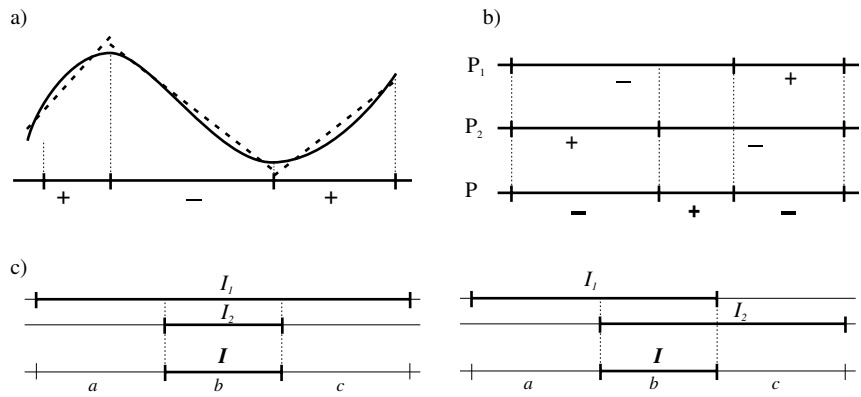}
\caption{Signed partition corresponding 
to a piecewise linear approximation
(panel a),
union of signed partitions (panel b),
and triplet $(a,b,c)$ for an interval
of a union of partitions.
see Sect. \ref{ssc}.}
\end{figure}
\clearpage

\newpage
\begin{figure}[p]
\centering \includegraphics{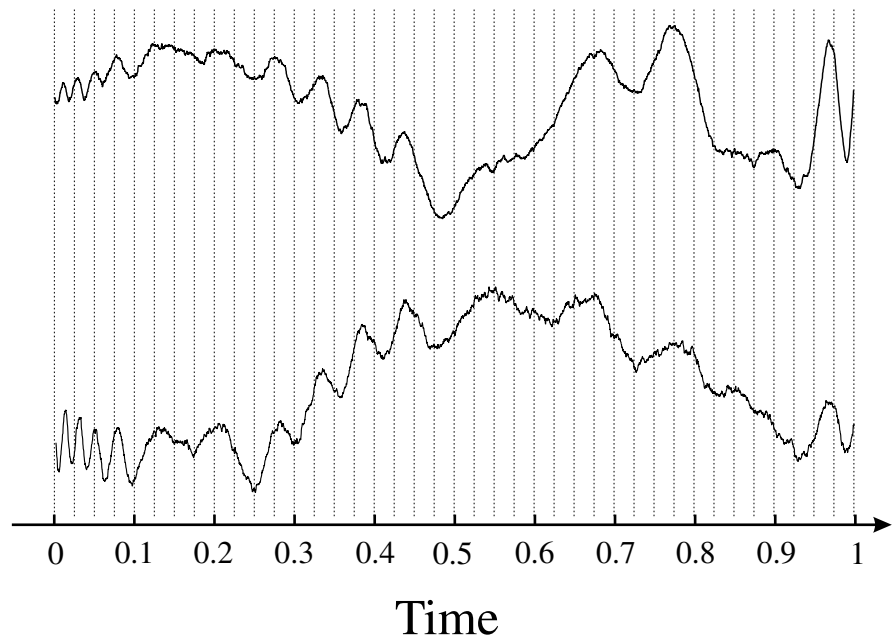}
\caption{Two signals coupled by an unobserved 
phenomenon.
The signals tend to change their
intermediate trends synchronously,
while their overall shapes are 
different.
The striking similarity is
observed at intervals $[0,0.1]$ and 
$[0.2,0.55]$.
Note also the common peaks at 
$t=0.675,0.775,0.975$.
See details in Sect. \ref{detect}.}
\end{figure}
\clearpage

\newpage
\begin{figure}[p]
\caption{Correlation (reciprocal distance) 
$\mu$ (\ref{dist2}) 
between two signals shown
in Fig. 15 (panel a) and two 
independent Brownian walks (panel b).}
\end{figure}
\clearpage

\newpage
\begin{figure}[p]
\centering\includegraphics{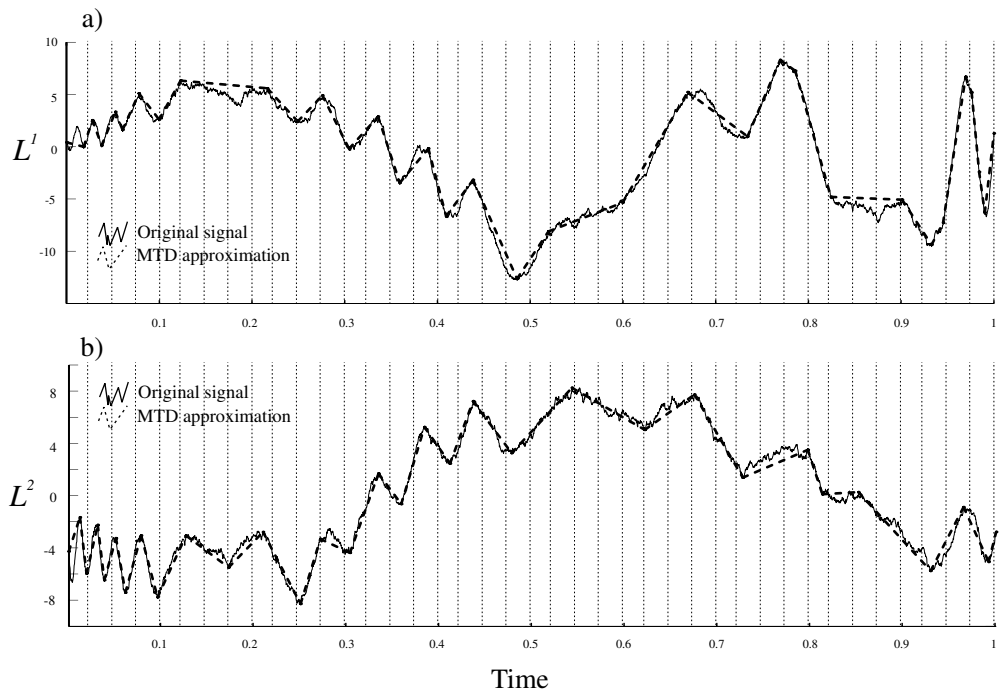}
\caption{Piecewise linear approximations 
$L^i,i=1,2$ of the signals $F_i(t)$ 
from Fig. 15 at the levels of maximal 
correlation $(l_1=15,l_2=14)$.
These approximations depict the 
intermediate-scale variations 
responsible for the signals' coupling.}
\end{figure}
\clearpage

\newpage
\begin{figure}[p]
\caption{Structure of the signals 
$F_i(t),i=1,2$ shown in Fig. 15.
a),e) Original signals $F_i(t)$.
b),f) Coupling parts $A_i(t)\cdot X(t)$.
c),g) Non-linear deterministic trends.
d),h) Random drifts.}
\end{figure}
\clearpage

\newpage
\begin{figure}[p]
\caption{Reconstruction (solid lines) of 
the coupling parts $A_i(t)\cdot X(t)$ 
(dashed lines).
See Sect. \ref{detect} for discussion.}
\end{figure}

\end{document}